# Generative AI Feedback, English Writing and Teacher Rubrics: A Multiple-Case Study of CyberScholar

[1]**Raigul Zheldibayeva,** [2]**Ana Karina de Oliveira Nascimento,** [3]**Vania Castro,** [3]**Bill Cope and** [3]**Mary Kalantzis,**

[1]Zhetysu University named after I. Zhansugurov; University of Illinois Urbana-Champaign - Bolashak International Scholarship, Kazakhstan
[2]Universidade Federal de Sergipe; University of Illinois Urbana-Champaign - Postdoctoral fellowship by the National Council for Scientific and Technological Development – CNPq, Brazil
[3]University of Illinois Urbana-Champaign

**Abstract:**

This multiple-case study examined the potential of a Generative AI (GenAI) tool, CyberScholar, to support K–12 students' writing across disciplines. This tool integrates teacher-provided rubrics, materials, and exemplars through Retrieval-Augmented Generation (RAG), producing criterion-specific formative feedback and ratings. The study involved 143 students and five teachers in grades 7 through 11 across five U.S. middle and high schools. Data sources included classroom observations, student post-surveys (n = 79), student focus group interviews (n = 18), and teacher surveys (n = 5). Qualitative analysis followed two cycles of coding to identify patterns within and across cases. Findings indicate that students valued CyberScholar's immediate, rubric-based feedback and noticed improvements in their writing as they revised, using it to refine organization, elaboration, and style. They also highlighted the tool's interactive, iterative qualities, which fostered revision and reduced reliance on teacher feedback. However, participants noted inconsistencies in the automated rating system and occasional misalignment with assignment expectations. Teachers reported that CyberScholar saved time on feedback and supported more targeted, higher-order instructional practices. The study underscores the promise of rubric-grounded GenAI formative feedback for developing writing skills, while emphasizing the need for human oversight, calibration of automated ratings, and attention to contextual factors shaping adoption.

# INTRODUCTION

The context for the research reported upon in this paper is discussions about Generative AI (GenAI) adoption, including its potential to provide immediate formative feedback to learners and reduce teacher workload. The paper does not aim to measure learning outcomes or establish causal effects. Instead, it aims to describe an implementation model, identify practical challenges and design decisions, and offer pedagogical recommendations grounded in classroom experience.

Writing has been considered an essential 21st-century skill for students to obtain college and employment readiness (Deane, 2022). At the same time, it has been a challenging competence for students to acquire. According to the Nation's Report Card (2022), only 27% of 12th-grade students demonstrated proficiency in writing on the National Assessment of Educational Progress.

Receiving detailed and well-balanced feedback that addresses both achievements and areas for improvement may help students develop writing skills and reinforce their learning. However, accessing in-depth formative feedback is still a challenge in many educational contexts. There are numerous factors that limit teachers from providing detailed feedback, including time constraints, as meaningful feedback takes time, large class sizes, quantity of materials for evaluation, and other responsibilities teachers have.

Recent studies have argued that Generative AI (GenAI) presents great promise for education, particularly for providing feedback (Cope and Kalantzis, 2019; Johnson, 2023; Mollick and Mollick, 2023). According to these authors, with thoughtful design and implementation, AI technologies could enhance formative feedback with continuous machine-mediated human assessment from multiple perspectives.

In this sense, this study proposes harnessing the potential of GenAI to support students' writing with formative feedback. This study introduces and examines the potential of CyberScholar, an assistant developed to support students' writing with formative feedback aligned with teachers' rubrics and instructional materials through fine-tuning. This is done through Retrieval

Augmented Generation (RAG) technology. i.e., materials are added to CyberScholar to supplement the information available in the underlying GenAI foundation model used in the platform. This ensures that feedback is aligned with educational standards, teacher expectations, and the rubric.

The adoption of feedback aligned to teachers' rubric is supported by research showing that rubrics significantly enhance transparency by providing clear expectations for students, offering consistent and objective evaluation for educators (Panadero et al., 2023). Stevens and Levi (2013) highlight the ways in which using rubrics to assess students' learning can provide focused, actionable feedback for students when it is adapted for many assignments and contexts.

Despite their importance, many scholars (Panadero and Jonsson, 2020; Brookhart, 2018; Zheldibayeva et al., 2025; Castro et al., 2026) acknowledge that there are few studies examining how students engage with rubrics while performing school tasks. In particular, there is a lack of research focusing on students' involvement with AI feedback connected with teachers' rubrics.

The present paper seeks to answer the following research questions: What is the potential of CyberScholar to provide formative feedback and support K–12 students' writing when AI feedback is aligned to teacher rubrics? What are the participants' roles and perspectives regarding the use of GenAI for feedback in student writing? The study involved 143 students from grades 7 to 11 and five teachers across five U.S. schools. This research examines the hypothesis that feedback to learners can be more valuable if calibrated through rubric-based and when GenAI has been integrated into instructional materials.

## Literature Review

This research is framed in the context of current debates about the impacts of GenAI on literacy. What makes GenAI particularly challenging is that it is a writing technology (Kalantzis & Cope, 2025a). It can undertake tasks traditionally assigned to students by teachers, such as writing a coherent text and composing it in response to a prompt. This has concerns about "cheating" at worst, or minimally "cognitive offloading" by students delegating some of their cognitive effort to AI (Gerlich, 2025). Our focus in this project was on the positive potential of GenAI in formative writing assessment, or how it may extend student writing and thinking-through-writing, rather than reduce it (Kalantzis & Cope, 2025b).

Prior empirical studies have explored how AI tools can support writing development. Luckin et al. (2016) have noted that tools such as Grammarly and QuillBot have the potential to enhance writing practice and provide individualized feedback on students' texts. Particularly, these AI tools, mostly from the era of Natural Language Processing (NLP) and preceding GenAI, have shown promise in assisting students to identify and correct grammatical mistakes more effectively (Godwin-Jones, 2022; Dembsey, 2017; Saeed, 2024). These studies center on writing mechanics, leaving unanswered questions about how these tools could function if they were integrated to instructional rubric to provide in-depth feedback in writing.

Other studies show the direct implication of GenAI in academic integration and performance. Qian (2025) conducted a systematic literature review to examine the initial academic response to GenAI and its adoption in higher education during 2023 and 2024. The study analyzed 262 peer review articles and identified that many themes from the pre-ChatGPT era persist, such as automated feedback, assessment, and learning support. However, new trends emerged, for example, fostering creativity, critical thinking, learning autonomy, and prompt literacy. This investigation has not identified studies that connect AI feedback with rubric alignment.

A study by Usdan et al. (2024) highlights that, with a customized educational curriculum and proper guidance, participants reduced writing time by 64.5% and improved writing quality from a B+ to an A. The findings show that GenAI can be a powerful tool for creating sophisticated content at the graduate level.

Van Niekerk et al. (2025) examined the GenAI-calibrated feedback offered to learners on their writing in higher education. The participants found the tool beneficial for enhancing the structure and flow of their writing, helping non-native English speakers achieve a more formal tone. Concerns were raised about the reliability of references, the lack of critical insight, and the potential objectivity of responses. These concerns align with Cope and Kalantzis (2023) who claim that a well-documented characteristic of AI is "algorithmic bias" along with racial/cultural, gender bias, the privacy of teachers and students; "hallucination" or the creation of false information; failure to acknowledge sources and to invent fictional sources; and the quality of the filters required to moderate these sources. In K-12 learning, all forms of AI raise a range of ethical concerns (Akgun & Greenhow, 2022). Qian (2025) also raises issues about AI overreliance, possibly leading to the outsourcing of critical cognitive and metacognitive skills of students.

Other studies have focused on GenAI for feedback in the K-12 settings. Lee and Moore (2024) and Shi and Aryadoust (2024) recognize that feedback generated by AI can offer timely and personalized feedback to learners. Wan and Chen (2023) evaluated feedback provided by GPT-3.5 on middle school science writing. The study showed that students tended to rate the feedback by humans and GPT equally on correctness, but they all rated the feedback by GPT as more useful. Similarly, Steiss et al. (2024) investigated the use of ChatGPT by 200 students from grades 6-12 in a U.S. school for offering formative feedback on students' drafts. The results showed that differences between ChatGPT and humans were modest when considering the overall quality of feedback and time-savings, but human feedback was better than ChatGPT feedback for ⅘ elements of formative feedback.

Fesler et al. (2026) elaborated a report on research findings about how AI affects teaching and learning. They found that although research on AI in K–12 has expanded, exhaustive causal evidence remains scarce. Only 20 high-quality studies were identified among more than 800 reviewed. Existing causal studies show that AI tools often boost student performance, but these gains frequently decrease when students are assessed without AI support. The evidence also highlights that tool design matters: AI systems that incorporate pedagogical guardrails, such as step-by-step reasoning instead of direct answers, tend to promote deeper learning. For educators, AI tools can reduce planning time and enhance instructional quality, especially for less experienced instructors. However, major gaps remain, especially regarding equity, student wellness, and impacts in U.S. K–12 contexts, as most causal studies occurred in different contexts. Overall, the report emphasizes the promising findings at the same time it stresses the need for more research to understand long-term learning effects, and the implications of AI use in schools.

Research specifically connecting GenAI feedback with teacher rubrics in the K-12 context is still limited. One of those studies, by Ekizoğlu and Demir (2025), explores the impact of AI feedback on the writing of 60 Turkish high school students in an EFL class. The experimental group of students used Grammarly to receive feedback on their drafts. They wrote multiple essays, which varied in genre and topic, including descriptive writing, opinion essays, and narrative writing. They received AI feedback according to four elements of the rubric: Content, Organization, Language Use, and Mechanics. Results indicated that students showed significantly greater improvement in overall writing performance compared to the control group (teacher-only feedback).

All these studies indicate the potential of GenAI tools to support feedback and writing development, while they also note limitations and ethical concerns, such as biases in AI-generated feedback and risks to student privacy, among others. The CyberScholar tool has been developed to apply its own filters, with built-in flagging for biases and other problems that may slip through the filter. The present paper introduces an additional dimension to the potential of GenAI for writing assistance by aligning AI-generated feedback with teachers' rubrics and classroom learning materials in K-12 contexts.

**Research Design**

This study adopted a qualitative research design grounded in an interpretivist paradigm (Erickson, 1986), aiming to explore the potential of CyberScholar to support K–12 students' writing across diverse subject areas by providing formative feedback aligned to teacher rubrics. Also, it focused on finding out the participants' roles and perspectives regarding the use of GenAI for feedback in student writing. Qualitative methods were chosen to capture the complexity of student interactions with GenAI, their experiences, and the contextual factors that shape writing development, along with formative feedback.

A multiple-case study design (Yin, 2018) was adopted, with each school serving as a bounded case. This choice was particularly suited to this research because it allowed for an examination of real-world classroom use of GenAI in writing. This approach enabled researchers to capture variation across urban and rural schools, middle and high school grade levels, and different subject areas. By emphasizing context and meaning-making, the qualitative design supported the study's goal of uncovering how GenAI might influence writing processes, engagement, and outcomes in authentic educational settings.

The focal workspace in CyberScholar is a multimodal editor with multilingual capacities. There, students can write and think-through-writing ("knowledge representation") with AI support in any subject area, Grade 4 and above. On the left of a vertically divided screen is the editor. On the right are tools for AI and human dialogue to support students' work. The interactions between both sides generate rich and granular source data both for assessment and research.

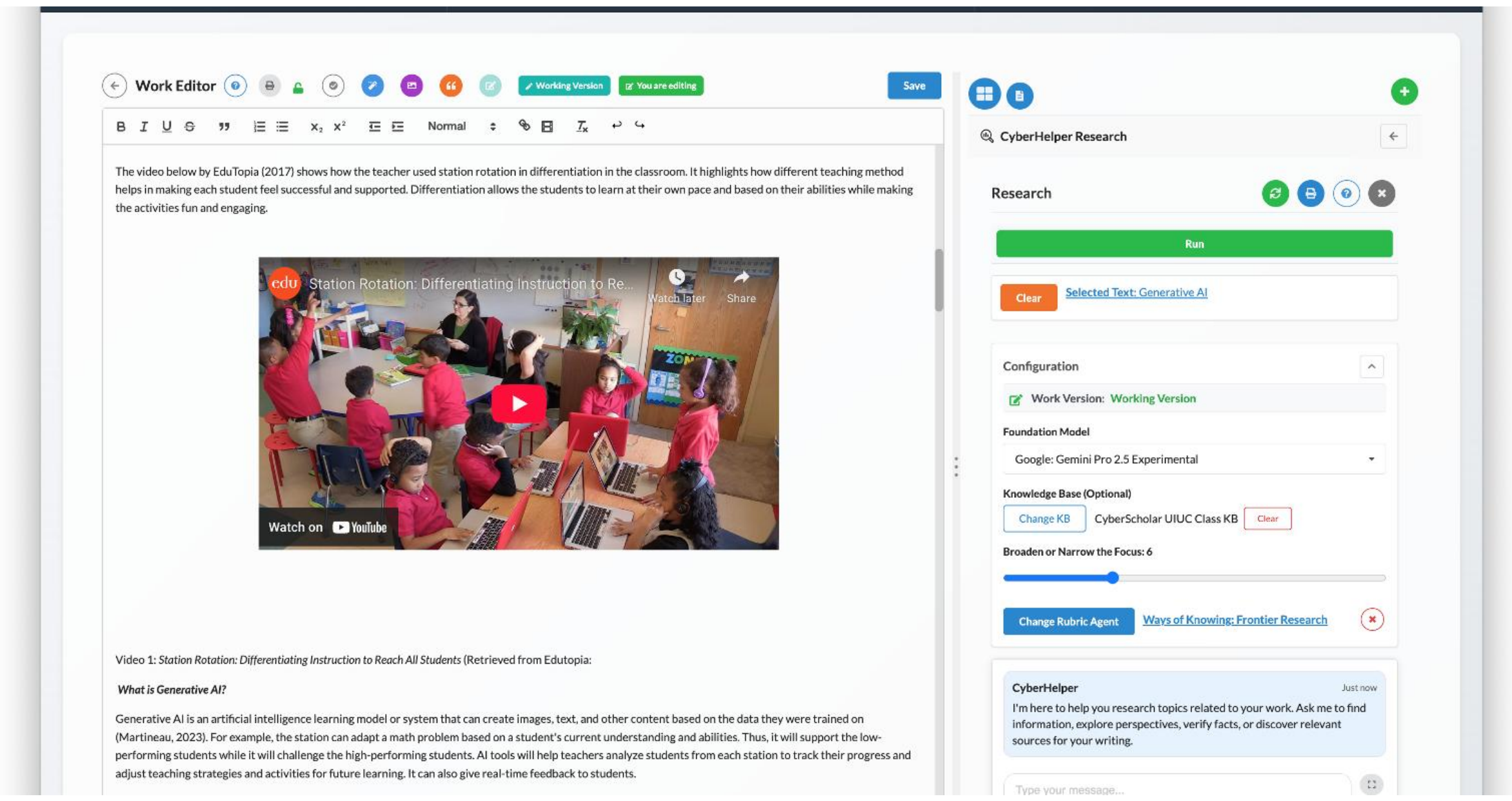

Figure 1. Work editor (Source: CyberScholar interface. Used with permission.)

Figure 2 below outlines the workflow and AI architecture of CyberScholar, which includes its available functions, detailed as follows. *CyberHelper* allows writers to request AI assistance during drafting by selecting text in the editor and activating the tool. *CyberReviewer* delivers quantitative feedback with qualitative narrative justification from AI rubric agents, peers, teachers, or self-reflection. *Composition Report* analyzes the extent and way students use AI to support their writing and thinking, including whether AI facilitates cognitive offloading or extends student thinking, by integrating generative AI with contextual logfile data such as keystrokes, timestamps, and clickstreams. The system is further supported by Rubric Agents, developed by teachers or instructional designers to align evaluation and feedback with disciplinary knowledge frameworks and expected learning standards through multiple-pass prompt engineering and chain-of-thought processes. The *Knowledge Base* consists of curated, educator-validated domain content stored in a vector database for retrieval-augmented generation. The User Profile enables the AI to respond in ways that support the learner's Zone of Proximal Knowledge. Foundation Models, which teachers may select from publicly accessible, open-weight, or commercial options, are integrated, while learner identities and work remain securely separated through CyberScholar's application programming interface.

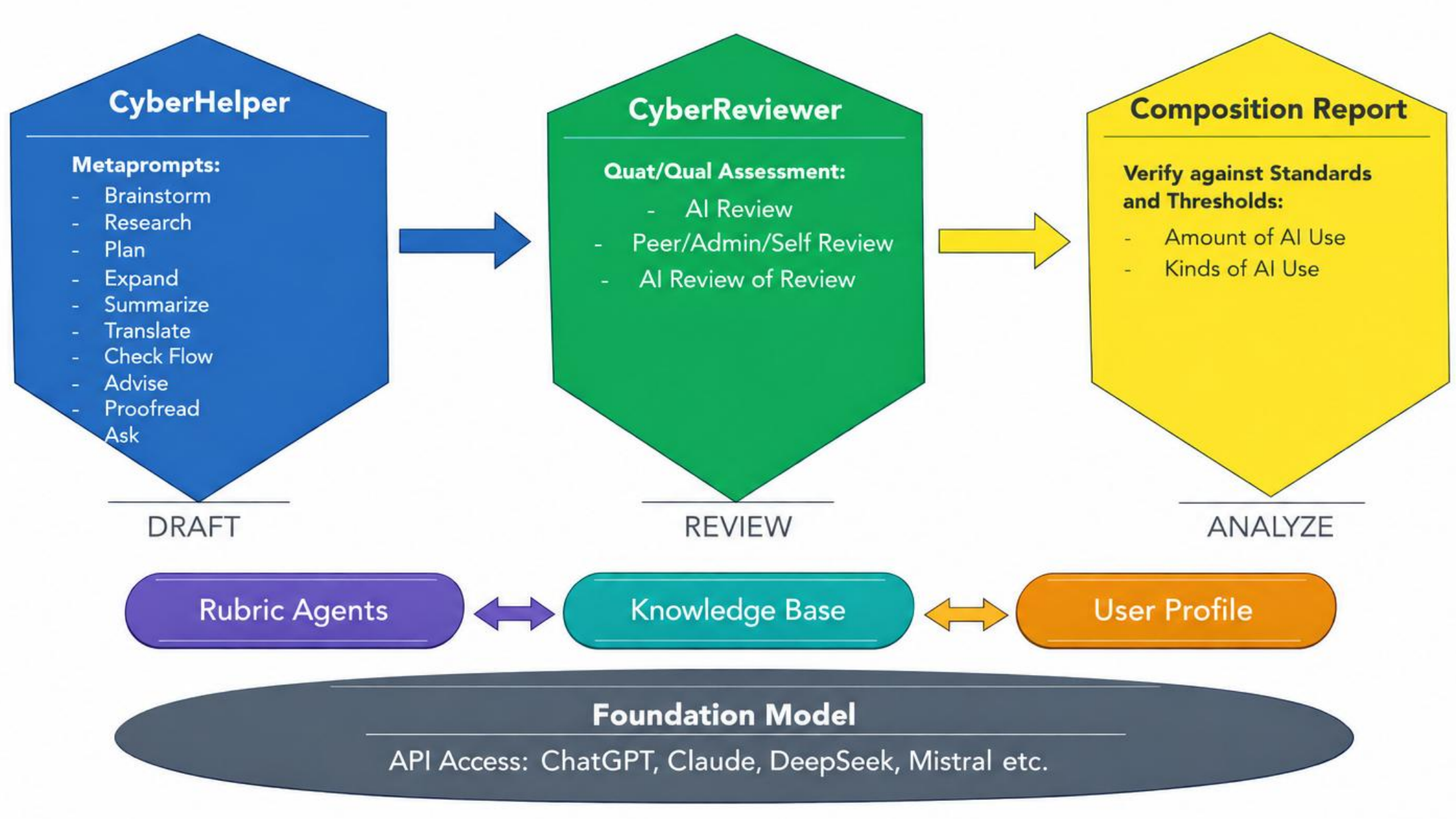

Figure 2. Workflow (Source: CyberScholar interface. Used with permission.)

The study focused on five distinct school settings where CyberScholar was piloted. Each case was defined by the school site (Schools A, B, C, D, and E), the participants (students and their teachers), the timeframe (one academic semester in 2024/2025), and the phenomenon under investigation: the use of CyberScholar to provide formative, rubric-aligned AI feedback to support K–12 writing.

This design enabled a rich, contextualized understanding of how the platform functioned across varied educational environments. This approach allowed comparative insights, and at the same time, preserved the depth of individual cases. For every school, the research team generated a structured analytic memo outlining student and teacher perceptions of AI feedback, participants' roles, challenges, behaviors, and classroom context. Analysis proceeded through a within-case examination of each school, followed by a cross-case analysis comparing patterns across cases to identify convergences and divergences in relation to the study's research questions.

Building on the qualitative multiple-case design, this research foregrounds responsible GenAI implementation and teacher professional learning as integral components of the design. In line with recommendations for designing for responsible GenAI use in K–12, the study relied on ethics and compliance protocols (e.g., separation of student identity from foundation models via API architecture) to shape concrete design decisions, including restricting data retention, disabling persistent model training on student submissions, and embedding bias-flagging and hallucination filters despite potential trade-offs in model flexibility and scoring stability. These decisions reflect a deliberate prioritization of privacy and equity over maximal automation. Additionally, the research design explicitly conceptualized teachers as co-implementers within a human-in-the-loop, cyber-social pedagogy: onboarding sessions emphasized rubric agent construction, interpretation of AI-generated feedback, calibration of ratings, and strategies for preventing overreliance or cognitive offloading, while also identifying areas where teachers required support (e.g., managing inconsistent star ratings, setting classroom norms for AI dialogue) versus areas of existing expertise (e.g., rubric design, formative assessment practices). By documenting these professional learning structures and ethical trade-offs as part of its methodological architecture, the study contributes not only as an exploration of student experience but also as a model for responsible, teacher-mediated GenAI integration in authentic K-12 contexts.

The trials were conducted in five public schools in the United States: four high schools and one middle school, which represent a diverse range of socioeconomic contexts, as it can be visualized below.

**Table 1.** Characteristics of Participating School Sites

| Site | School Type / Location | Enrollment | Grades | Student Demographics | Grade Distribution | Student – Teacher Ratio | Dropout Rate | Socioeconomic Context |
|---|---|---|---|---|---|---|---|---|
| A | Suburban high school, Midwest | 824 | 9–12 | 35% White, 30% Hispanic/Latinx, 24% Black/African American | Grade 9 (233), Grade 10 (235), Grade 11 (184), Grade 12 (175) | 13:1 | 2.8% | 99.6% qualified as low-income based on eligibility for free/reduced-price lunch, substitute care status, or public aid |
| B | Suburban university laboratory high school, Midwest | 314 | 8–12 | 35% Asian, 35% White, 14% Hispanic/Latinx, 7% Black/African American | Grade 8 (64), Grade 9 (64), Grade 10 (60), Grade 11 (65), Grade 12 (61) | 9:1 | 0% | Data on low-income/FRPL eligibility were redacted and not reported by the state |
| C | Rural middle school, Northwestern | 69 | 6–8 | 36.5% White, 63.5% students of color, including 58.7% American Indian or Alaska Native and 4.8% identifying as two or more races | Grade 6 (23), Grade 7 (21), Grade 8 (19) | ~15:1 | Not reported at school level | 37.5% classified as low-income/economically disadvantaged |
| D | Suburban high school, Midwest | 1,825 | 9–12 | 45.5% White, 26.8% Hispanic/Latinx, 20.3% Black/African American | Grade 9 (474), Grade 10 (468), Grade 11 (427), Grade 12 (456) | 20:1 | 2.5% | 63.3% classified as low-income based on eligibility for free/reduced-price lunch or similar state criteria |
| E | Centralia High School | 869 | 9–12 | 69.6% White, 7% Hispanic/Latinx, 11.5% Black/African American, 10.4% two or more races | Grade 9 (220), Grade 10 (222), Grade 11 (236), Grade 12 (191) | 15:1 | 9.8% | 63.4% classified as low-income based on eligibility for free/reduced-price lunch or similar state criteria |

These five cases reflect varied geographic, socioeconomic, and cultural contexts. This diversity enabled us to examine CyberScholar's implementation across different grade bands, subject areas (English Language Arts - ELA, Social Studies, History), and community settings (suburban, rural). The five contexts, due to their diverse nature, are relevant for cross-case analysis, as reported in this paper.

## Participants

Although 143 students participated in the study and were part of classroom observations, 79 completed all stages of the study until completion of the post-survey, 18 participated in the focus group interview phase, and the 5 teachers completed the post-survey. The description of participants from each site is presented next:

**Table 2. Characteristics of Participants**

| School | Teacher characteristics | Teacher experience with GenAI | Number of participating students | School years |
|---|---|---|---|---|
| **School A** | One ELA teacher for juniors and seniors; 18 years of teaching experience (6 years in subject) | Very experienced with digital technology; used Grammarly and ChatGPT; very comfortable integrating AI | 6 (3 completed post-survey; 3 participated in focus group) | Grade 11 |
| **School B** | One 10th-grade Modern World History teacher; 16 years of experience (8 years in subject) | Very experienced with digital tools; used ChatGPT; somewhat comfortable integrating AI | 48 (11 completed post-survey; 4 participated in focus group) | Grade 10 |
| **School C** | One Language Arts teacher (grades 7–8); 17 years of experience (3 years in these grades) | Very experienced; used School AI and Diffit; very comfortable with AI integration | 17 (8 completed post-survey; 5 participated in focus group) | Grades 7–8 |
| **School D** | One Social Studies teacher (grades 11–12); 20 years of experience (14 years in subject) | Very experienced; used MagicSchool.ai; somewhat comfortable integrating AI | 37 (30 completed post-survey; 0 focus group) | Grades 11–12 |
| **School E** | One ELA teacher; ~10 years of experience | Experienced with digital tools; used MagicSchool AI; moderate AI integration | 35 (27 completed post-survey; 6 participated in focus group) | Grade 10 |

## Data Collection

The research procedure followed a clearly defined, ethically approved sequence comprising recruitment, consent, training, classroom implementation, and post-implementation data collection. All activities were conducted in alignment with the University of Illinois Urbana-Champaign Institutional Review Board protocol (IRB24-0912) and district-level Student

Online Personal Protection Act (SOPPA) agreements, ensuring that data protection, privacy, and participant rights were upheld throughout the study.

Recruitment and consent were conducted in 2024 and 2025. Schools were invited to participate through existing institutional partnerships, and five school sites agreed to integrate the CyberScholar into an upcoming writing unit. Teachers who volunteered for the study provided informed consent, after which parental permission and student assent were obtained for all minors in the participating classes. After completion of consent procedures, students and teacher pre-surveys were administered to establish an understanding of participants' prior experience with AI, writing challenges, and expectations regarding the tool to be used.

Following recruitment, teacher and student training were conducted which, depending on the school site, occurred from September to November 2024. Classroom implementation took place from November 2024 to March 2025 as part of each teacher's regular instructional sequence. Students drafted their written assignments on the CyberScholar platform and requested AI feedback aligned with the teacher's rubric. They revised their work in response to this feedback and resubmitted improved drafts. Teachers monitored students' use of the tool, clarified instructional goals, and supported students in interpreting and applying AI-generated formative feedback.

Observations were conducted by multiple researchers using field notes, which were later discussed collaboratively and then consolidated. Observation notes included: (a) the number of students present, (b) indicators of student motivation and engagement, and (c) the types of questions students asked while interacting with the tool. These elements provided insight into both behavioral engagement and the instructional scaffolding required at each site. In addition, the research team conducted support check-ins with participating teachers on an as-needed basis via email and Zoom meetings.

In the final stage, post-implementation data collection for Schools A-D occurred in December 2024 and March 2025 for School E. Some students completed an online post-survey (n=79) consisting of open-ended questions about their experiences with CyberScholar, and all teachers (n=5) completed a similar reflective post-survey. These were administered after the writing unit and designed to capture participants' evaluations of the experience, the tool, perceived benefits and limitations, and reflections on its influence on writing or teaching practices.

Sample items from the post-survey for students included closed and open-ended questions such as: “Did the AI Helper tool make writing assignments easier for you?”, “Can you provide examples of how the AI Helper helped your writing?”. Sample items from the post-survey for teachers also included closed and open-ended questions, such as: “To what extent did the AI Helper support writing instruction in your classroom?”, “How did the tool impact students’ writing skills?”, and “What challenges or improvements do you identify when using the AI Helper for writing instruction?”.

Semi-structured focus group interviews were then conducted with a subset of students (n=18) to gain deeper insight into how the tool influenced writing, learning processes, and classroom dynamics. Students' focus group interview questions invited participants to describe how they engaged with AI formative feedback during drafting and revision, how they decided whether to accept or reject suggestions, and how they perceived the role of CyberScholar within their broader classroom experience. Focus group interviews were video-recorded, transcribed using Otter.ai, later revised, and finally supplemented with the researchers' field notes.

Across all phases, the procedures adopted clearly distinguished research activities (surveys, focus group interviews, and classroom observations) from tool development activities. All collected qualitative data were included in the analysis and are reported in the Findings section. While the CyberScholar platform continued its broader development cycle during the academic year, no development meetings, internal testing logs, or engineering decisions were included in this study. The procedure reported here reflects the implementation of the tool in authentic classroom settings and the systematic collection of qualitative data to understand participant experiences.

**Data Analysis**

In order to observe the findings, codes were initially attributed to the data by three of the research team members, the first listed as coauthor of this paper. The investigators met to collaboratively review the data generated through each instrument: post-surveys, classroom observations, and focus group interviews. Each time the investigators met, they analyzed a different set of data from a particular instrument, examining it for recurring patterns. The coding approach was inductive, with codes emerging directly from the data. The research team developed a collaborative codebook, where they registered the initial codes inductively from the data, capturing recurring patterns, behaviors, and perceptions observed in the post-surveys, focus group interviews, and classroom observations. Therefore, deductive codes were also informed by the research questions and the pre-existing literature on assisted writing and formative feedback by AI. Then, the codebook was iteratively refined: codes were defined with clear descriptions and examples, merged when overlapping, and added when new patterns emerged.

For the analysis, each code represented a concise term or phrase, developed by the research team, capturing a core idea or significant theme within a segment of linguistic data. Coding assigned interpretive meaning to individual units, facilitating pattern identification, organization into categories, and analytical interpretation (Saldaña 2013). Whenever a disagreement occurred about themes or codes, a new round of analysis was carried out so that an agreement involving the three coders could be reached.

Triangulation was conducted across data sources. After initial and second-cycle coding of each dataset, the research team compared emerging codes and themes across the data instruments. Themes were kept when supported by evidence from multiple data sources. To ensure transparency and avoid conflicts of interest, no members of the research team were involved in developing CyberScholar, and all coding and analysis were conducted independently from the platform's developer.

In order to analyze the data, following Saldaña (2013), initial coding was developed. It involved segmenting qualitative data into distinct units for detailed examination, enabling the identification of patterns through comparative analysis of similarities and differences. This methodological approach is broadly applicable across the spectrum of qualitative research designs, and that was chosen for this particular study.

The second-cycle pattern coding involved pattern coding, which is about the recognition of data segments that share similar initial codes and the systematic organization of these segments into thematic groupings. This technique was selected for its suitability in second-cycle coding, particularly in facilitating the emergence of overarching themes and supporting the interpretive pursuit of underlying explanations within the dataset (Saldaña, 2013).

**Findings**

The results of the thematic analysis of the participants' responses across the instruments used revealed that CyberScholar has the potential to support K-12 students' writing by providing rubric-aligned formative feedback. According to students' and teachers' perspectives regarding the use of GenAI for feedback, the support for writing was achieved through: 1. The Provision of Detailed and Specific feedback; 2. The Recognition of Writing Improvement Through AI Feedback; 3. The Delivery of Ratings Connected to Rubric Criteria; 4. The Interactive Nature of the GenAI Tool.

Of the respondents to the post-survey and focus group, most students explicitly stated that they would like to use the CyberScholar tool again for writing assignments. The same was shared by the participating teachers. A minority of students across the five schools indicated in the post-survey that they were hesitant about using the tool again. All teachers stated that they would recommend it to other teachers. Initially knowing most participants would approve of the GenAI tool to support writing assignments, the next step was to work on the first cycle coding and then the second cycle coding. Based on the researchers' analysis and following pattern coding, the following themes emerged.

**The Provision of Detailed and Specific Feedback**

This theme highlights aspects that participants praised. Evidence for this theme emerged consistently in students' and teachers' post-survey responses and was reinforced during focus group interviews. The majority of students across the five schools explicitly mentioned in their comments positive opinions about the in-depth and individual feedback. Students noted that this allowed them to identify specific areas for revision, which contributed to their overall writing development. Some students highlighted that the tool provided concrete ways to fix issues, helping them understand their mistakes and how to improve their writing, especially in areas in need of more nuanced and sophisticated expression.

One student highlighted that the AI review corrected her text in chronological order, which made it easy for her to make corrections. Another student appreciated the tool's organization and the use of basic bullet points. A third student noted that the AI feedback was focused and made it easy to locate specific information or areas needing improvement, particularly in the revision process. The teachers highlighted the advantage of having GenAI connected to their teaching, thus providing feedback aligned to their requests. The following quotes are examples of these views across the five sites, since this theme emerged across the whole dataset:

"Both kinds of feedback [my teacher's and CyberScholar's] were very straightforward, and told me directly what needed to be improved" (Student 57 - School E)

“It did allow for feedback tailored to the student's specific writing.” (Teacher 01 - School A)

Some students decided to run the AI review again after their revisions. They noticed improvements in their scores, which meant that the assistant tool had provided them with valuable formative feedback to enhance their writing development. In addition, participating students realized that the feedback was aligned with what the teacher had specifically required. For example, during classroom observation, it was possible to visualize that Student 2 - School A improved the following items from the rubric “Compare” and “Contrast” score from 0 to 2. After the first round of CyberScholar feedback, some students were able to identify some similarities and differences between the first and the last revised texts, which resulted in a higher rating and improved writing. It was found that the formative feedback aligned with teacher rubrics from the AI tool was useful for enhancing their writing.

Data from focus group interviews reinforced that students perceived CyberScholar’s feedback as detailed and specific. And this was considered central to them. Participants from School E emphasized that CyberScholar not only identified issues in their writing but also pointed to “specific places” that needed revision in the text and offered concrete suggestions for improvement. Students highlighted that the level of precision, especially regarding grammar, sentence structure, and organization, helped them to locate and correct errors, contributing to a more productive revision process.

Beyond students’ experiences, teachers noted that detailed, rubric-aligned AI feedback reduced the time spent on repetitive comments, allowing them to focus on higher-order concerns such as argument development, coherence, and student reasoning. As one teacher explained in the post-survey:

“Cyberscholar allowed for an additional opportunity for formative assessment that I would not have had time for otherwise. I appreciate being able to have the students submit to the AI assistant before I review a draft. It seems to save some time and effort on my part”. (Teacher 02 - School B)

Although students valued the specificity of AI feedback, some teachers expressed concern that very specific suggestions could potentially limit students’ opportunities for critical thinking, analysis, and intellectual discovery. One teacher cautioned in the post-survey:

“We need to make sure that the AI is allowing student analysis and creative thinking. There are times where it goes too far and suggests specific topics that the students should investigate… The students still need to do the heavy intellectual lifting.” (Teacher 01- School A)

Another teacher recognized the benefits of tailored feedback but contrasted it with the contextual personalization favored by human instruction and interaction:

“It did allow for feedback tailored to the student's specific writing. But it was missing the personalization I can provide when I give feedback. For example, I might say to a student, "This is an improvement from your elaboration last time! Great job!" or "Remember in class when we talked about [specific topic redacted]? Try that here." (Teacher 03 - School C)

These findings demonstrate different perspectives among participants. On one hand, students praised the detailed and specific feedback; on the other, teachers emphasized the importance of preserving opportunities for critical reflection, contextual encouragement, and independent meaning-making.

**The Recognition of Writing Improvement Through AI Feedback**

This theme highlights how students described becoming aware of improvements in their writing through their interactions with CyberScholar, particularly on how they interpreted those suggestions as evidence of growth. Across the five schools, teachers recognized advancement in most of their students' writing. The majority of students reported that the tool helped them recognize areas for improvement in their writing. They mentioned how it provided recommendations for their texts, based on teachers' expectations and criteria of evaluation as described in their rubrics. Students described learning about advanced writing practices, such as strengthening conclusions, improving coherence between ideas, and word choices, and provided recommendations for sentence structure and clarity. The following quotes demonstrate what students noticed in their writing improvements:

"We learned that our conclusions should not just simply restate what we were talking but, especially in a History assignment, we should also try to explain the significance of this project." (Student 11 - School B)

"It helped me clear up some sentences that sounded a bit too clunky and helped me rewrite them." (Student 13 - School D)

Focus group interviews conducted in School E reinforced that students recognized clear improvements in their writing as a result of interacting with the feedback provided by CybserScholar. Several participants noted that the tool helped them "go back and find the places" that needed correction, indicating an increasing awareness of their own writing weaknesses. Also, the feedback pointed out how to address the writing problems. Some students, even though mentioning occasional redundancy in the feedback received, consistently described the tool as "really helpful" and "one of the best things" they had used to check their writing assignments. Participants' voices suggest that the feedback contributed to a sense of progress and helped them increase confidence in their writing process.

Teachers also acknowledged that the AI feedback was effective; however, they highlighted that its impact depended heavily on students' willingness and readiness to engage in revision and improve their writing skills. One teacher described how students who were motivated to improve their writing engaged deeply with the AI feedback:

"For those students looking to genuinely improve in their writing, [CyberScholar] was of great help. [...] Specifically, I had a student who used language like 'a reader might see' or other tentative-sounding phrases. After receiving comments, he went back to sound more definitive and confident in his assertions." (Teacher 05 – School E)

The same teacher highlighted how convergence between AI and teacher feedback increased students' acceptance of revision:

"Another student who struggled with transitions told me he didn't think his writing sounded abrupt, but after the AI Helper told him the same thing, he stated it must be an issue for both the technology and the teacher to comment on the same issue. So he worked to correct it." (Teacher 05 – School E)

Similarly, other teachers emphasized that students who actively engaged with the feedback benefited most, while others struggled to act on it. One reason, as noted by Teacher 04 - School D, was that there is "a group of students who are so unaware of AI and how to utilize it that they didn't know what to do with the feedback." Another reason, identified by Teacher 02 - School B, was that some students liked the feedback but were too reluctant to put in the effort required to make changes.

**The Delivery of Ratings Connected to Rubric Criteria**

This theme highlights participants' satisfaction with the ratings provided by CyberScholar and how the tool applied rubric criteria to their texts. Across all sites, students praised the feedback connection to teachers' expectations, which was expressed by focusing on the rubric. Teachers also recognized the value of having feedback aligned to their expectations. However, the delivery of ratings revealed divided opinions among the participants as to how to visualize them.

Some students pointed out that the tool provided them with structured and consistent feedback by assigning ratings and scores based on evaluation criteria created by their teachers. The tool's alignment with the teacher's rubric was seen as positive and understood as a big difference compared to their individual uses of GenAI in real life. Therefore, students appreciated that the AI explicitly referenced rubric categories like "writing mechanics." This suggests the tool made the assessment criteria more visible and understandable, which is often a challenge in traditional feedback. Besides that, the way the tool was programmed, automated feedback mimicked teacher expectations, bridging the gap between abstract rubric language and concrete writing performance. This reflects a shift from generic feedback, often common in uses of GenAI tools, to precision-guided revision, which is a hallmark of effective formative assessment, the aim of the tool being tested.

However, the way the ratings were displayed in numbers of stars generated differing opinions. For example, a student explained in a focus group interview that the ratings helped him understand how much to improve transitions and accurately reflected the quality of his writing, making the stars a true representation of his journal entries. Similarly, another student appreciated the stars as a rating system because they motivated her to earn more stars in the next round of AI feedback than she had previously received. During the focus group interviews, students from School B mentioned how they appreciated the star-based grading system for providing a clear visual understanding of their results. One of the students mentioned that this encouraged her to work even harder. The following quotes are examples of these views in the student post-survey:

"I liked the star ratings and how in-depth the explanations were" (Student 02 - School A)

“Yes, I liked how it rated my essay in each category, so I could see where I needed to focus my attention, which was nice. This way I could see the improvement of my writing each time I submitted a revised version.” (Student 26 - School D)

On the other hand, some participants, all from School D, found the ratings inconsistent, noting that they sometimes received different scores for the same submission without making any changes.

"I did not like that when I put in the same paper twice, it gave me two very different rankings. It made me unsure of what score my paper should have actually gotten and made things more difficult." (Student 35 - School D)

"The ratings were kinda random, like I told the AI to reevaluate and it gave me a better score even though I didn't make any changes." (Student 33 - School D)

This finding might signal a greater familiarity of those students with GenAI, since this was the only site in which participants freely tested submitting the same text for AI review without making any changes. In addition to that, the inconsistency in ratings also relates to issues of AI reliability, as mentioned earlier in this paper, something that should always be considered by teachers when using it pedagogically.

Teacher 03 - School C noted that the star ratings were beneficial because students viewed the stars more as feedback than as grades.

“The students were encouraged when they made revisions and saw an improvement in their score. They trusted the results and didn't see it as a grade but rather feedback. When I score their drafts, they see it as a grade, and some feel like they can settle for their first draft score. The use of stars instead of a letter grade or number works great to motivate them without thinking about the grade.”

However, Teacher 02 - School B had reported that the 4-star rating was too simplistic and could be distracting: “My students are already too distracted by the stars and not paying close enough attention to the feedback that the assistant gives.” He suggested replacing the stars with points (0-10) or percentages. He referred to the stars as "basic" and explained that the scoring should be tougher for his students. Moreover, he suggested that having both stars and feedback on their work might be confusing for his students, explaining that if his high-performing students received 3 out of 4 stars, they might feel satisfied and not read the feedback for any particular criterion.

**The Interactive Nature of the GenAI tool**

This theme presents students' appreciation for the interactive feature of the CyberScholar tool, which allowed them to engage in a dynamic conversation with the tool during the revision process. Only a few students across the schools explicitly praised the interactivity of the GenAI tool. It is important to note that this functionality was added as the research progressed and was available only for Schools C, D, and E. Students asked GenAI follow-up questions for further clarification, making the formative feedback process more engaging and useful. This functionality enabled students to be more active in the learning process rather than passively receiving comments. This can be seen in the following quote:

“I like how it responds to my questions. Especially when I asked, "Can you further elaborate what I need to fix in this section?" it gave me good feedback and explained thoroughly when I asked it to elaborate further.” (Student 17 - School D)

"I liked that we could communicate with the AI and ask it questions. For example, I always asked for a summary." (Student 67 - School E)

Data from School E highlighted that students actively engaged with CyberScholar in an interactive, back-and-forth way. Thus, the tool was treated not merely as a feedback provider but as a conversational partner in the revision process. Several participants described asking the AI follow-up questions for different reasons, such as when they did not understand a comment, to request synonyms, to ask for title suggestions, or clarification about how to strengthen specific sentences. Students also reported using the tool iteratively, that is, after revising a paragraph, they would ask AI "how it looked". Participants' views demonstrate that the interactive feature supported a dynamic cycle of feedback, revision, and reassessment.

The focus group interview also revealed that students valued the tool's responsiveness and adaptability, noting that it provided targeted guidance, especially when students interacted with it. According to some participants, AI feedback looked different from their teacher's due to how the feedback was provided. According to them, teachers' feedback was typically broader, less individualized, while AI offered concrete examples, pointed to exact sentences, and explained how to fix them when asked. This ability to question the tool directly and receive immediate, specific responses seemed to enhance students' sense of agency during revision, reinforcing the perception that CyberScholar worked as an interactive writing support available to students.

In addition to student post-survey responses and focus group interviews, classroom observation notes from School C provided converging evidence of this interactive engagement. During live implementation, one student was observed using the conversational box beneath the criteria-based feedback to ask follow-up questions, which the student appeared to find particularly helpful for understanding how to revise his writing. The research team also observed that another student appreciated that the tool allowed her to interact independently, get vocabulary suggestions, and receive organized feedback with bullet points.
Teacher 04 - School D explicitly highlighted this functionality as a key strength, mentioning that:

"The students really liked being able to ask for more details, to distill something to its essence, or to ask for more examples or summarize."

Teacher 05 - School E also mentioned students' use of the interactivity as a positive aspect in the construction of students' understanding of the feedback, and then their revised writing:

“I had several students who meticulously went through the feedback they received, asked me questions, asked AI questions, and worked to improve their drafts.”

This finding suggests that when interactivity was available and visible to participants, it promoted deeper engagement with AI feedback. In contrast, the teacher from School C did not comment on this feature, which may indicate either limited student use or that other aspects of the tool were more prominent in that context.

**Other less common themes**

Other less common themes emerged. According to the data, CyberScholar also has the potential to support K-12 students' writing through: the understanding of it as a resource for self-directed learning, the provision of personalized and quick feedback, the possibility of saving students' time, as the tool quickly identified areas for improvement in students' texts, and provided fast feedback.

## Discussion

This study found that students and teachers across all five sites valued CyberScholar’s rubric-aligned, detailed feedback and interactive dialogue with AI, particularly as a formative support during the revision process. Students mentioned that the tool helped them identify concrete areas for improvement and understand how to revise rather than simply what was wrong. However, participants questioned the reliability of automated ratings, in particular when scores appeared inconsistent across submissions. Teachers also raised concerns about the ratings, especially because they could limit opportunities for student agency, critical thinking, and contextualized encouragement.

These findings highlight both the promise and limitations of GenAI in supporting writing development in K-12 contexts. Overall, CyberScholar demonstrated potential to enhance students' writing by connecting AI feedback directly to teacher rubrics, while also raising important questions about consistency and trust in automated scoring.

Compared to prior studies, this work extends the literature on GenAI writing tools in meaningful ways. Previous research has emphasized the role of tools such as Grammarly and QuillBot in improving mechanics and grammar (Godwin-Jones, 2022; Dembsey, 2017; Saeed, 2024), and highlighted benefits in higher education contexts (van Niekerk et al., 2025; Usdan et al., 2024). However, these studies did not explore how AI feedback could be aligned with instructional rubrics. By integrating teachers' rubrics and materials, CyberScholar moves beyond generic feedback to provide guidance that is pedagogically grounded and directly connected to classroom expectations. This alignment represents a significant change from earlier work. While prior investigations have largely concentrated on higher education (Qian, 2025; van Niekerk et al., 2025), this research demonstrates how GenAI can be adapted to K–12 contexts, where formative feedback is central to learning. The multi-site approach helps capture diverse student and teacher perspectives across different school settings, including variations in familiarity with GenAI. In addition to that, students' possibility of revising their work after receiving feedback and resubmitting it to see improvements is something that can be implemented in school practices, which is quite difficult to put into practice with limited teacher time.

Theoretically, the findings contribute to the literature on formative assessment by showing how timeliness, specificity, and student agency can be improved through rubric-aligned AI feedback. Students reported that CyberScholar made rubric criteria more visible and actionable, bridging the gap between abstract evaluation language and concrete writing performance. This exemplifies how AI can mediate the relationship between teacher expectations and student self-regulation. Rather than replacing human feedback, CyberScholar functioned as a formative partner, enabling students to engage more actively in their own learning process. At the same time, the inconsistencies in ratings draw attention to the need for theoretical caution: AI-mediated practices must be critically examined for reliability.

In practical terms, the implications of this study are varied and significant. For teachers, CyberScholar demonstrates how AI can assist their job of providing feedback, allowing them to have more time for more complex instructional interactions, while still requiring human work to ensure accuracy and contextual appropriateness. For schools, the findings emphasize the importance of aligning AI tools with existing rubrics and curriculum, rather than adopting generic systems that may not reflect local pedagogical priorities. For students, the tool offers opportunities to develop agency and writing improvement, but also raises risks of overreliance on AI. Ethical and equity considerations remain vital: data privacy protections (e.g., FERPA, SOPPA) must be implemented, and inconsistent scoring may compromise fairness. Finally, while CyberScholar has the potential to support underprivileged schools by providing timely feedback, the digital divide persists, reminding educators that equitable access to technology is a condition for these advantages to be realized.

The opportunities and challenges of implementing formative GenAI feedback in K–12 schools, aligned with teachers' rubrics to support students' writing, highlight the need to critically assess its role in classroom practice.

**Limitations**

As a qualitative multiple-case practice study focused on feasibility and participant perceptions, this research has several limitations that delimit the claims it can make. Methodologically, the sample was relatively small and uneven across sites (with only 79 students completing post-surveys and 18 participating in focus groups), participation was voluntary, and the implementation spanned a single semester, limiting generalizability and raising the possibility of self-selection and novelty effects. The study relied primarily on self-reported perceptions, observational field notes, and thematic coding rather than independent, standardized measures of writing quality, inter-rater reliability statistics, or longitudinal performance data. Consequently, this practice paper cannot determine whether CyberScholar caused measurable improvements in writing achievement, whether gains were sustained over time, or how AI-supported revision compares to well-designed teacher-only or peer-feedback conditions in controlled designs. It also cannot establish the reliability, validity, or fairness of automated ratings across demographic groups, nor can it fully assess issues such as cognitive offloading, shifts in metacognitive strategy use, or differential impacts on novice versus advanced writers. What remains unknown includes how rubric-aligned GenAI feedback functions across broader grade levels and disciplines, how teachers' varying levels of AI literacy influence outcomes, how students internalize feedback over extended periods, and how contextual AI logfile data might correlate with demonstrated learning gains. Future research would therefore require mixed-method and experimental or quasi-experimental designs, larger and more diverse samples, validated writing assessments with blind human scoring, longitudinal tracking of revision trajectories, analysis of AI–human feedback convergence, and equity-focused studies examining bias, access, and differential effects. Such research would move beyond feasibility and perception to establish causal impact, reliability, and scalable models for responsible GenAI integration in K–12 writing instruction.

## Conclusions

This study has examined CyberScholar as a bounded case of rubric-aligned GenAI feedback in authentic K–12 classrooms rather than as a validated intervention, and our conclusions are therefore necessarily modest. The findings indicate that rubric-anchored, iterative AI feedback can function as a structured formative prompt that makes assessment criteria more visible and actionable during drafting and revision. However, the evidence presented here supports claims about perceived usefulness and patterns of engagement rather than demonstrated improvements in writing quality. What this study shows is that, under certain conditions, students and teachers experienced rubric-aligned GenAI feedback as supportive of revision processes and time management, while also encountering misalignment, over-specific suggestions, and variability that required human judgment. Rather than validating CyberScholar as an instructional solution, this research contributes an exploratory account of how implementation of GenAI-mediated formative feedback was interpreted, negotiated, and scaffolded in diverse school contexts.

Nevertheless, the study demonstrates that a GenAI formative assessment tool such as CyberScholar has the potential to enrich formative assessment in K–12 writing by aligning AI-generated feedback with teacher rubrics and curricular expectations. Students valued the detailed, rubric-based guidance, which supported both their perception of writing improvement and their agency in the revision process. At the same time, concerns about rating reliability and ethical considerations remind us that human supervision and equitable access remain crucial. By situating GenAI feedback in K-12 classroom sites, this research contributes to a more pedagogically grounded understanding of AI in education.

Several pedagogical lessons emerge from this intervention. First, what worked well was the explicit alignment of AI feedback to teacher-authored rubrics, which supported clearer connections between criteria and revision decisions. Educators considering similar tools should prioritize tight curricular integration over generic AI deployment. Second, what surprised the research team was the degree to which students selectively accepted, rejected, or questioned AI suggestions. This was an observable indicator of agency that became visible only when teachers framed feedback as advisory rather than authoritative. Third, what did not work as expected included the star-based rating system, which in some contexts distracted from qualitative commentary or created misplaced confidence. Future implementations should recalibrate or decenter automated scores and foreground narrative justification. Finally, the study underscores that rubric-aligned GenAI feedback makes pedagogical sense only when embedded in structured instructional routines, supported by professional development, and accompanied by explicit norms for critical evaluation of AI output. In the absence of such conditions, and in circumstances where teacher mediation, institutional policy guidance, and infrastructural supports are weak, AI feedback may be counterproductive, narrowing inquiry or encouraging superficial compliance. Thus, the implications of this work are conditional and context-sensitive. GenAI can extend formative assessment practices, but only when treated as a mediated component of a broader pedagogical ecology grounded in teacher expertise, critical interrogation, and sustained institutional support.

**Funding**

This research was funded by the Science Committee of the Ministry of Science and Higher Education of the Republic of Kazakhstan under the project "Artificial intelligence as an assistant in acquiring academic English by non-language major students: Randomized study" (AP25794537).

Zheldibayeva, R., Nascimento, A. K. D. O., Castro, V., Kalantzis, M., & Cope, B. (2025). The impact of AI-driven tools on student writing development: A case study. *Online Journal of Communication and Media Technologies, 15*(3), e202526. https://doi.org/10.30935/ojcmt/16738